\documentclass[12pt,a4paper,%final
preprint]{iopart}

\usepackage{iopams}  
\usepackage{graphicx}
\usepackage[breaklinks=true,colorlinks=true,linkcolor=blue,urlcolor=blue,citecolor=blue]{hyperref}

\begin{document}

\renewcommand{\L}{{\cal L}}
\renewcommand{\l}{\mathfrak l}

\begin{flushright}
UTHEP-672 \\
RIKEN-TH-209 \\
RIKEN-QHP-186\\
\end{flushright}
\title[Infinite circumference limit of CFT]{Infinite circumference limit of conformal field theory}

\author{Nobuyuki  Ishibashi%$^{1,2}$
}
\address{%$^1$
Faculty of Pure and Applied Sciences, University of Tsukuba, \\
Tsukuba, Ibaraki 305-8571, Japan }
%\address{$^2$Address Two, Neverland}
\ead{ishibash@het.ph.tsukuba.ac.jp}

\author[cor1]{Tsukasa Tada}
\address{RIKEN Nishina Center for Accelerator-based Science, \\
Wako, Saitama 351-0198, Japan}
\ead{\mailto{tada@riken.jp}}
 
\begin{abstract}
We argue that an infinite circumference limit can be obtained in 2-dimensional conformal field theory by adopting $L_0-(L_1+L_{-1})/2$ as a Hamiltonian instead of $L_0$. The theory obtained has a circumference of infinite length and hence exhibits a continuous and heavily degenerated spectrum as well as the continuous Virasoro algebra. The choice of this Hamiltonian was inspired partly by the so-called sine-square deformation, which is found in the study of a certain class of quantum statistical systems. The enigmatic behavior of sine-square deformed systems such as the sharing of their vacuum states with the closed boundary systems can be understood by the appearance of an infinite circumference.
\end{abstract}


\bigskip\bigskip
\begin{center}
\textit{Version published as  \\ Journal of Physics A: Mathematical and Theoretical  {\bf 48} (2015) 315402}.\\
DOI: \href{http://dx.doi.org/10.1088/1751-8113/48/31/315402}{10.1088/1751-8113/48/31/315402}\ {}\footnote{Content from this work may be used under the terms of the Creative Commons
Attribution 3.0 licence. Any further distribution of this work must maintain attribution to the author(s) and the title of the work, journal citation and DOI.}
\end{center}
\maketitle

Among all Virasoro generators $L_n$ that play  essential roles in the analysis of a conformal field theory (CFT), three of them, $L_0$, $L_1$ and $L_{-1}$, form a subalgebra that is isomorphic to $sl(2,\mathbb{R})$ and corresponds to the global conformal transformation.  The Casimir operator of this subalgebra can be expressed as
\begin{equation}
C_2=L_0^2-L_+^2-L_-^2,
\end{equation}
if we introduce the notation as follows;
\begin{equation}
L_+=\frac{L_1+L_{-1}}{2} \ ,\ L_-=\frac{L_1-L_{-1}}{2i}.
\end{equation}

In analogy to the (2+1)-dimensional Lorentz transformation, the space spanned by $L_0$, $L_+$, and $L_{-}$ is apparently divided into three distinctive regions. The first region is  "time like" and contains $L_0$ and small perturbations around it. Any vector within this region can be transformed to $L_0$ up to some numerical factor, by the global conformal transformation, $SL(2,\mathbb{R})$. This is actually the region that we would have in mind when we demand the invariance of the vacuum on the basis of the physical equivalence of the states connected by the global conformal transformation. The second region is "space like" and contains the linear combination of $L_+$ and $L_{-}$.  The region between these two is the last one and  is called the ``light like'' region. This region is represented by either $L_0-L_+$ or $L_0-L_-$.

If we further evoke  the analogy with the Lorentz geometry, the "time like" region corresponds to a "massive" representation. Because we observe the spectrum of $L_0$ in this region, the "mass" in this case should be the inverse of the circumference, or the finite scale of a CFT  \cite{Belavin:1984vu,Cardy:1984rp}. Thus, it is natural to induce that the "light like" region corresponds to the "massless" representation, having an infinite circumference. In this study, we argue that if we consider the generator in the ``light like'' region, say $L_0-L_+$ (plus the anti-holomorphic part ${\bar L}_0-{\bar L}_+$, to be exact), as a Hamiltonian, we can obtain a CFT with an infinite circumference. One might say that the following analysis is the ``light-cone quantization'' of a CFT in terms of the global conformal $SL(2,\mathbb{R})$.

\begin{figure}[htbp]
\begin{center}
\includegraphics[clip,width=8cm]{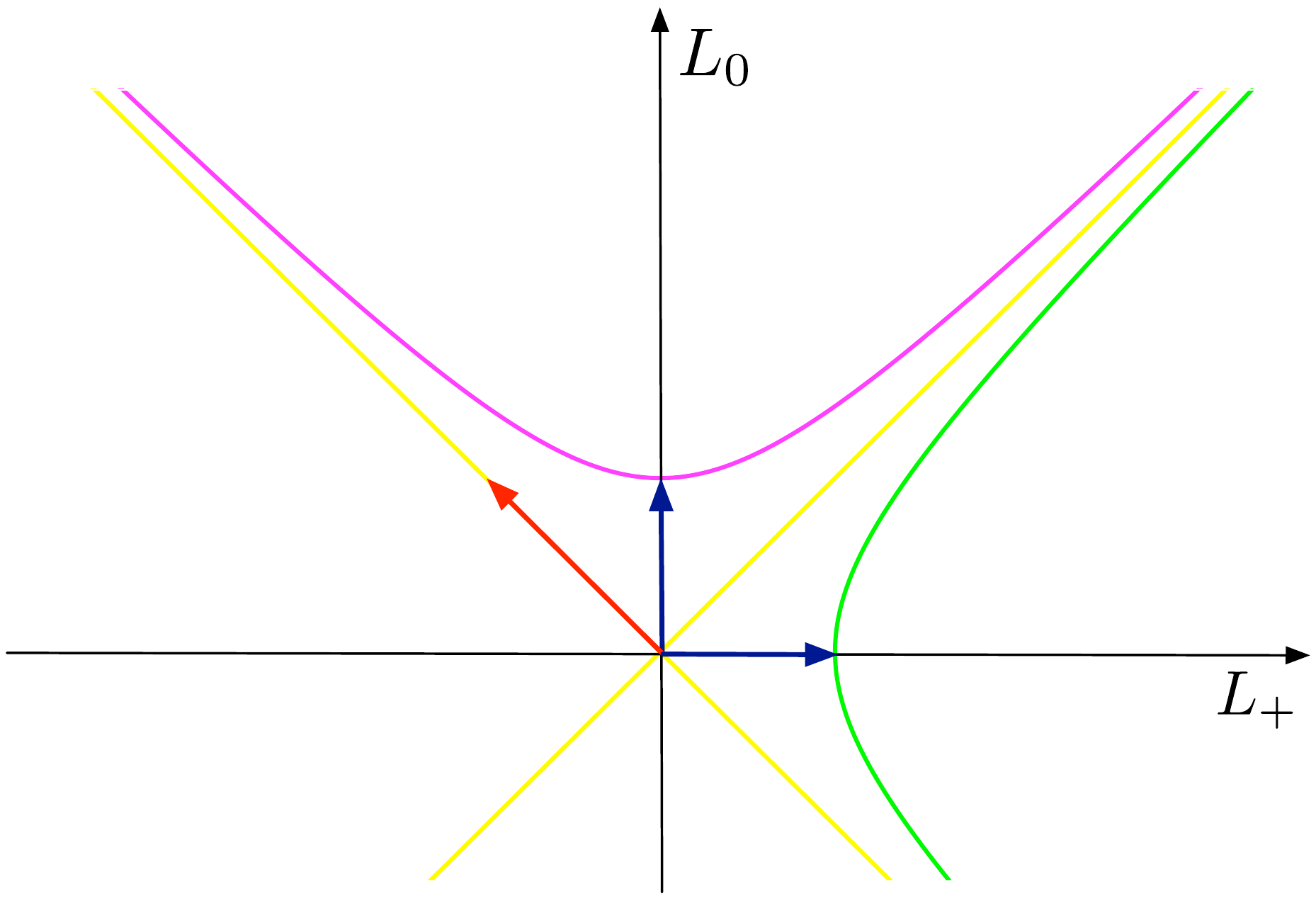}
\caption{
Two vectors that are $SL(2,\mathbb{R})$-equivalent to the cyan and green hyperbolae respectively, are shown along with the red arrow that represent $L_0-L_+$. 
%There are three distinct region which, in analogy to Lorentz geometry, we label as ``time like,'' ``light like,'' and ``space like,'' respectively.
%
%Three vectors, which are $SL(2,\mathbb{R})$-equivalent to the solid curve, the dotted line and the dashed curve respectively, represent three distinct region.
}
\label{fig1}
\end{center}
\end{figure}

Let us start with the generator that corresponds to $L_0-L_+$, given by
\begin{equation}
-z\partial_z+\frac{z^2+1}{2}\partial_z=\frac12 (z-1)^2\partial_z, \label{eqn:l0def}
\end{equation}
in the complex coordinate. We denote the above generator as $\l_0$. The following analysis may exhibit a partial similarity with those in \cite{Kiermaier:2007jg,Rastelli:2001hh}.
Let us introduce the eigenfunctions of $\l_{0}$ as follows:
\begin{equation}
\l_{0} f_{\kappa} (z)= -\kappa f_{\kappa}(z). \label{eqn:fkdef}
\end{equation}
%Ishibashi-san's suggestion 03/31/2015
%The notation above may appear a little awkward because of  the negative sign in front of the eigenvalue, but  w
We can readily solve Eq. (\ref{eqn:fkdef}) as
\begin{equation}
f_{\kappa}(z)=A_{\kappa}e^{\kappa\int^{z}\frac{2d%z}{(z-1)^{2}}}  : by the suggestion of the referee 05/29/2015
w}{(w-1)^{2}}}=A_{\kappa}e^{-\frac{2\kappa}{(z-1)}}, \label{eqn:fksol}
\end{equation}
where $A_{\kappa}$ denotes a $\kappa$ dependent constant.

Using $f_{\kappa}(z)$, we can introduce a set of  differential operators as follows;
\begin{equation}
\l_{\kappa}=\frac12 (z-1)^2f_{\kappa}(z)%\frac{\partial}{\partial z} : by the suggestion of the referee 05/29/2015
{\partial_z}. \label{eqn:lkdef}
\end{equation}
Note that the above definition includes the case for $\l_{0}$ (\ref{eqn:l0def}), as its $\kappa=0$ case, provided that $A_{0}=1.$ Therefore, we take $A_{0}$ to be unity in the equations given below.

The commutation relation among the $\l_{\kappa}$ generators is  easily calculated to be
\begin{equation}
[\l_{\kappa}, \l_{\kappa'}]=(\kappa -\kappa')\frac12 (z-1)^2f_{\kappa}(z)f_{\kappa'}(z)%\frac{\partial}{\partial z} : by the suggestion of the referee 05/29/2015
{\partial_z}\ .
\end{equation}
Noting that
\begin{equation}
f_{\kappa}(z)f_{\kappa'}(z)%=A_{\kappa}A_{\kappa'}e^{(\kappa+\kappa')\int^{z}\frac{dz}{g}} : by the suggestion of the referee 05/29/2015
=\frac{A_{\kappa}A_{\kappa'}}{A_{\kappa+\kappa'}}f_{\kappa+\kappa'}(z)
\end{equation}
from the formal solution of $f_{\kappa}(z)$ (\ref{eqn:fksol}), we arrive at the Witt algebra or classical Virasoro algebra;
\begin{equation}
[\l_{\kappa}, \l_{\kappa'}]=(\kappa -\kappa')\l_{\kappa+\kappa'}, \label{eqn:Witt}
\end{equation}
if we impose the condition that $A_{\kappa}A_{\kappa'}=A_{\kappa+\kappa'}$, which is satisfied by 
\begin{equation}
A_{k}=e^{{const.}\kappa}.
\end{equation}

We have introduced the eigenfunctions of $\l_{0}$, $f_{\kappa}(z)$ as a method of defining $\l_{\kappa}$, but let us now calculate the action of $\l_{\kappa}$ on $f_{\kappa}(z)$.
It is easy to derive that
\begin{equation}
\l_{\kappa}f_{\kappa'}(z)=f_{\kappa}(z)\l_{0}f_{\kappa'}(z)=-\kappa'f_{\kappa}(z)f_{\kappa'}(z)=-\kappa'f_{\kappa+\kappa'}(z).
\end{equation}
The action of $\l_{\kappa}$ on $f_{\kappa'}(z)$ shifts the eigenvalue of $f_{\kappa'}(z)$ in the amount of $\kappa$ yielding $f_{\kappa+\kappa'}(z)$, as well as multiplies it by  $-\kappa'$. Therefore,  the Witt algebra (\ref{eqn:Witt}) can be represented over the (Hilbert) space spanned by $f_{\kappa}(z)$'s.

The argument presented above can be duplicated for another set of differential operators starting from
\begin{equation}
{\bar \l}_{0}\equiv\frac12({\bar z}-1)^{2}%\frac{\partial}{\partial {\bar z}}: by the suggestion of the referee 05/29/2015
\partial_{\bar z}, \label{eqn:L0bardef}
\end{equation}
where ${\bar z}$ denotes the complex conjugate of $z$. Thus we have constructed two independent sets of the classical Virasoro algebra.

Now we would like to consider the time-development of the system driven by the generator $\l_{0}+{\bar \l}_{0}$. If we parametrize the time by $t$ and space by $s$, we obtain the following set of equations:
\begin{eqnarray}
&&-\frac{\partial}{\partial t} \equiv \l_{0} +{\bar  \l}_{0} ,
\\
&&-\frac{\partial}{\partial s} \equiv i \left( \l_{0} -{\bar  \l}_{0}\right).
\end{eqnarray}
It solves in a consice form;
\begin{equation}
t+is=\frac{2}{z-1}. \label{eqn:tandstoz}
\end{equation}
If we use the Cartesian coordinates $z=x+iy$, where
\begin{equation}
\left\{\begin{array}{c}x=1+\frac{2t}{t^2+s^2} \\  y=\frac{2s}{t^2+s^2}\end{array}\right. ,
\end{equation}
it follows readily that
\begin{equation}
\left(x-1-\frac{1}{t} \right)^{2}+y^{2}=\frac{1}{t^{2}}.
\end{equation}
Therefore, the trajectory of the constant $t$ is a circle with a radius $\frac{1}{|t|}$ whose center is located at $z=1+\frac{1}{t}$. The significance of this contour is that it always passes through the point $z=1$. Note that as the contour starts from $z=1$ and encircles back to $z=1$, the parameter $s$ ranges from $-\infty$ to $\infty$.

If we rewrite the time translational vector field in the Cartesian coordinates, we obtain
\begin{equation}
\frac{\partial}{\partial t}=\frac{-(x-1)^{2}+y^{2}}{2}\frac{\partial}{\partial x}-(x-1)y\frac{\partial}{\partial y}.
\end{equation}
This yields a dipole field located at $z=1$ in accordance with the previously mentioned fact that the constant time contour always passes through $z=1$. While the conventional treatment \cite{Belavin:1984vu} leads to the so called  radial quantization \cite{Fubini:1972mf}, %a word ``termed'' added in proof 07/08/2015
termed %
after the configuration of the time-translational vector field, the present situation may be called as the ''dipolar quantization.''
\begin{figure}[htbp]
\begin{center}
\includegraphics[clip,width=8.5cm]{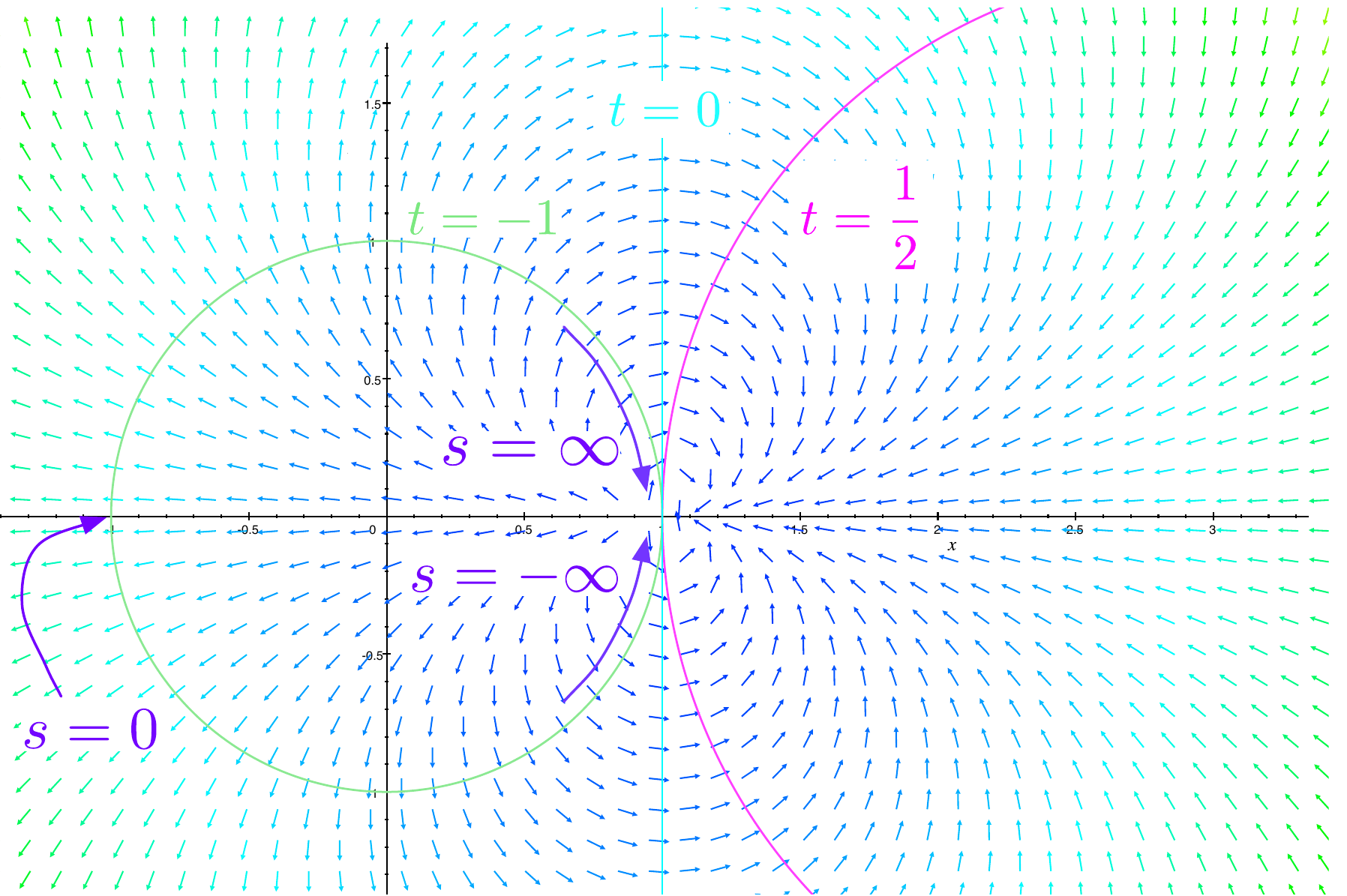}
\caption{Equal time contours for $t=-1$, $0$ and $\frac12$ are shown on the ``dipolar'' vector field generated by $\l_{0}+{\bar \l}_{0}$. The parameter $s$ assumes the value $0$ on the real axis other than at $z=1$, where $s$ becomes either $\infty$ or $-\infty$ depending on the way to approach.}

\label{fig1}
\end{center}
\end{figure}

Once we understand the constant time contour, we can define the  conserved charges as follows;
\begin{equation}
\L_{\kappa} \equiv \frac{1}{2\pi i}\oint^{t=const.} dz (-\frac12 (z-1)^{2}) e^{\frac{2\kappa}{z-1}}T(z), \label{eq:calLkappadef}
\end{equation}
where $T(z)=T_{zz}(z)$ is the energy momentum tensor of the original CFT. Note that for $\kappa =0$
\begin{equation}
\L_{0}=\frac{1}{2\pi i}\oint^{t=const.} dz (-\frac12 (z-1)^{2})T(z)=L_{0}-\frac{L_{1}+L_{-1}}{2}. \label{eqn:calL0def}
\end{equation}

We can further calculate the commutation relations among the charges defined above using the operator product expansion of the energy momentum tensor
\begin{equation}
T(z)T(w) \sim \frac{c/2}{(z-w)^{4}}+\frac{2T(w)}{(z-w)^{2}} + \frac{\partial_{w}T(w)}{z-w} + \cdots .
\end{equation}
The result reads
\begin{equation}
[\L_{\kappa} , \L_{\kappa'} ]=(\kappa - \kappa')\L_{\kappa+\kappa'}
+\frac{c}{12}\kappa^{3}\delta(\kappa+\kappa'). \label{eqn:ViraLkappa}
\end{equation}
We have now obtained the continuous Virasoro algebra with central charge $c$ %
\footnote{The continuous Virasoro algebra was also recently considered in Ref. \cite{Doyon:2013paa}.
 }%
%Suggested by Ishibashi-san 03/23/2015
\footnote{Modifying the definition of $\L_{k}$ with an extra $\frac{c}{24}\delta(\kappa)$ included, will yield the more familiar expression of the central charge, $\frac{c}{12}(\kappa^{3}-\kappa)\delta(\kappa+\kappa')$. The added extra term can be attributed
to the Schwarzian derivative for the transformation: $z \rightarrow e^{\frac{2}{z-1}}$.
}
 .
When deriving the %central charge in :by the suggestion of the referee 05/29/2015
second term in Eq. (\ref{eqn:ViraLkappa}) 
which corresponds to the central extension%addition 05/29/2015
,  it is probably the easiest to consider the $t=0$ contour, which is a line that connects $z=1$ with the infinity:
\begin{equation}
z=1+i\frac{2}{s}, \ -\infty \leq s\leq\infty.
\end{equation}
Then, the term in question becomes
\begin{equation}
\frac{c}{12}\int^{\infty}_{-\infty}\frac{ds}{2\pi}\kappa^{3}e^{is(\kappa+\kappa')}=\frac{c}{12}\kappa^{3}\delta(\kappa+\kappa').
\end{equation}

Now let us exploit the algebraic structure of Eq. (\ref{eqn:ViraLkappa}) to obtain information about its spectrum. If we denote an eigenstate of $\L_{0}$ with an eigenvalue $\alpha$ with an additional index $\sigma$ denoting possible degeneracy
\begin{equation}
|\alpha , \sigma \rangle \ , 
\end{equation}
as
\begin{equation}
 \L_{0}|\alpha , \sigma \rangle=\alpha |\alpha , \sigma \rangle ,
\end{equation}
operating $\L_{\kappa}$ on $|\alpha , \sigma \rangle$ yields
\begin{equation}
\L_{\kappa}|\alpha , \sigma \rangle =|\alpha-\kappa , \sigma \rangle
\end{equation}
based on  commutation relation (\ref{eqn:ViraLkappa}). Therefore, starting from the vacuum or any other energy eigenstate, we can construct an eigenstate for $\L_{0}$ with an arbitrary eigenvalue because $\kappa$ can assume any real value. 

%%%
%Insertion to respond to referee's comment
At this juncture, it would be beneficial if we had a similar equation to, as  the case of ordinary 2d CFTs,  
\begin{equation}
L_{n}|0\rangle=0  \ \  \hbox{for} \  n \geq -1 ,  \label{eqn:Ln0eq0}
\end{equation}
which, in due course, ensures that the energy spectrum of 2d CFT is bounded below %by 06/08/2015
at $|0\rangle$.  Recall that the argument	%lead by Ishibashi-san 07/09/2015
leading to Eq. (\ref{eqn:Ln0eq0})
stems from the requirement of the regularity of the product of the vacuum and the energy-momentum tensor which is placed at $t = -\infty \ (z=0)$, 
\begin{equation}
\lim_{t\rightarrow-\infty}T(z)|0\rangle=\lim_{z\rightarrow0}\sum L_{n} z^{-n-2}|0\rangle. \label{eqn:Txvac}
\end{equation}
Were it not for Eq. (\ref{eqn:Ln0eq0}), the above expression (\ref{eqn:Txvac}) would have been divergent. Let us examine the suitable variant of (\ref{eqn:Txvac}) for our case.
The reciprocal expression of (\ref{eq:calLkappadef}) turns out to be
\begin{equation}
T(z)=\int_{-\infty}^{\infty} d\kappa \frac{4}{ (z-1)^{4}}e^{-\frac{2\kappa}{z-1}}\L_{\kappa} ,
\end{equation}
and incorporating (\ref{eqn:tandstoz}) yields
\begin{equation}
\lim_{t\rightarrow-\infty }T(z)|0\rangle=\lim_{t\rightarrow-\infty }\int_{-\infty}^{\infty}  d\kappa \frac{ (t+is)^{4}}{4}e^{-\kappa(t+is)}\L_{\kappa}|0\rangle.
\end{equation}
Divergence arises from $e^{-\kappa t}$ as we take $t$ to be $-\infty$ for any positive $\kappa$, hence we were led to 
\begin{equation}
\L_{\kappa}|0\rangle=0 \  \hbox{for}\  \kappa >0, \label{eqn:calLkappapluszero}
\end{equation}
in order that the above expression is regular. Eq%.by the suggestion of the referee 06/02/2015
uation (\ref{eqn:calLkappapluszero}) suggests that the spectrum of the system is bounded below by $|0\rangle$, at least for the states that can be constructed by the multiplication of $\L_{\kappa}$.

%%%
%The original paragraph below
%
%On the other hand, the integrand in Eq. (\ref{eqn:calL0def}) can be expressed in terms of the integral of $\L_{\kappa}$:
%\begin{equation}
%-\frac12(z-1)^{2}T(z)=-\int_{-\infty}^{\infty}d\kappa \frac{2}{(z-1)^{2}}e^{-\frac{2\kappa}{z-1}}\L_{\kappa}.
%\end{equation}
%If we assume the regularity of the above operator when it acts on the vacuum state $|0\rangle$ while considering the  $t\rightarrow-\infty$ limit, we must  demand that
%\begin{equation}
%\L_{\kappa}|0\rangle=0 \  \hbox{for}\  \kappa >0, \label{eqn:calLkappapluszero}
%\end{equation}
%as  $L_{n}|0\rangle=0 $ is usually required for $n \geq -1$ in the analysis of ordinary 2d CFTs. Eq%.by the suggestion of the referee 06/02/2015
%uation (\ref{eqn:calLkappapluszero}) suggests that the spectrum of the system is bounded below by $|0\rangle$, at least for the states that can be constructed by the multiplication of $\L_{\kappa}$.

Thus, we have established that the theory  with the Hamiltonian given by
\begin{equation}
H=\L_{0}+{\bar \L}_{0}
\end{equation}
exhibits a continuous spectrum. This is consistent with the argument at the beginning of the present correspondence, as well as the previously noted observation that the variable $s$ which is conjugate to the momentum, takes values from $-\infty$ to $\infty$.

%<< Conclusion
%
%SSD

This observation offers an elucidation of  the phenomenon known as the sine-square deformation (SSD)  \cite{SSD}, at least for the case involving a CFT \cite{Katsura:2011ss}. It was found in \cite{SSD} that a certain class of quantum systems exhibit %ed :by the suggestion of the referee 06/02/2015
identical vacua regardless of whether closed- or open- boundary conditions were imposed, if the coupling constants of the open-boundary system are %were: by the suggestion of the referee 06/02/2015
modulated as 
\begin{equation}
g\sin^{2}(%2 : by the suggestion of the referee 06/02/2015
\pi\frac{x}{L}).
\end{equation}
Here $g$ is the original coupling constant and $L$ is the size of the system. Note that the coupling constant becomes zero at the both ends $x=0$ and $x=L$ as it should be for the open-boundary condition, while it assumes its original strength $g$ just at the middle of the system $x=\frac{L}{2}$. If SSD is applied %for : by the suggestion of the referee 06/03/2015
to 2-dimensional conformal field theories \cite{Katsura:2011ss}, it is easy to see that the resulting Hamiltonian \cite{Katsura:2011ss} becomes just
\begin{equation}
L_{0}-\frac{L_{1}+L_{-1}}{2} +{\bar L}_{0}-\frac{{\bar L}_{1}+{\bar L}_{-1}}{2} .
\end{equation}
SSD for a 2d CFT produces exactly the same  Hamiltonian discussed in the present study. The implications of this fact for string theory were previously discussed by one of the present authors in \cite{Tada:2014jps,Tada:2014kza}. It had been somewhat enigmatic that these two systems with different boundary conditions share the same vacuum state because intuitively, the lowest energy state should be the most affected by the global structure of the system. Now, we know that  SSD systems possess continuum spectra implying that the system has an infinitely large space. The distinction between the open- and closed-conditions at the ends that located infinitely far away becomes irrelevant. It would be interesting to see if this explanation is applicable for other SSD systems besides CFT considered here.
As a matter of fact, it was shown in \cite{Hotta:2014} that  the first %
few %: by the suggestion of the referee 06/03/2015
excitation
energies %: by the suggestion of the referee 06/03/2015
 of several SSD fermionic systems on finite lattice %correspond : by the suggestion of the referee 06/03/2015
are proportional to the inverse-square of the system size, hence hinting that this is indeed the case.
%

%mention to \phi(1)
Another point worth mentioning in regard of the relation between the present analysis and SSD is the significance of the point $z=1$. There found \cite{Tada:2014kza} new solutions for %SSD: by the suggestion of the referee 06/03/2015
the states that are annihilated by $L_0-L_+$, by one of the present authors:
$e^{L_{-1}}\phi(0)|0\rangle$, where $\phi$ is a primary field,
and by H. Katsura:
$\sum_{n>1} L_{-n}|0\rangle$.
These solutions can be recast as
\begin{equation}
e^{L_{-1}}\phi(0)|0\rangle=\phi(1)|0\rangle,
\end{equation}
and 
\begin{equation}
\sum_{n>1} L_{-n}|0\rangle=T(1)|0\rangle,
\end{equation}
respectively, thus signify the %the :deleted 06/03/2015
point $z=1$. This resonates the peculiar role played by the point $z=1$ in the dipolar quantization  explained earlier.

The structure of the Hilbert space of the proposed theory could be very complex. 
It is apparent that for any given eigenvalue, there are many ways to multiply $L_{\kappa}$; hence,  the eigenstates are likely to be degenerated in general. Nonetheless, a set of vectors expressed as
\begin{equation}
|\kappa\rangle_{0} \equiv \L_{-\kappa}|0\rangle, \kappa >0,
\end{equation}
offers an orthonormal set of vectors because
\begin{equation}
{}_{0}\langle \kappa' |\kappa\rangle_{0} =\langle 0| \L_{%-    typo corrected 04/21/15
\kappa'}\L_{-\kappa}|0\rangle=\langle 0|[ \L_{%-   typo corrected 04/21/15
\kappa'},\L_{-\kappa}]|0\rangle
\end{equation}
amounts  to the equation given below owing to the commutation relation (\ref{eqn:ViraLkappa}):
\begin{equation}
\frac{c}{12}\kappa^{3}\delta(\kappa'-\kappa)\langle 0 |0\rangle, %  by suggestion of the referee 06/03/15
\end{equation}
if we assume the Hermite conjugate takes the form
\begin{equation}
\L_{\kappa}^{\dagger}=\L_{-\kappa}. \label{eqn:HermiteconjugateL}
\end{equation}
% Proposal 05/28/2015 in response to the second referee. 
While the above assumption (\ref{eqn:HermiteconjugateL}) seems natural and also implies  the Hermiticity of $\L_{0}$, it should be subject to further investigation, which will be affirmatively addressed in our future publication \cite{IshibashiTada2b} .

A straightforward generalization of the present analysis would also be worth considering:
\begin{equation}
{\cal H}=L_{0}-\frac{L_{n}+L_{-n}}{2} +\hbox{anti-holomorphic part},
\end{equation}
where $n$ is an integer larger than unity. It  is easy to see that these systems involve  disconnected spaces. The detailed analysis for these generalizations %
%Correction for JPA
% is
%left for future study.
 as well as more extensive and thorough exposition of the results found in this note, are left for future publication \cite{IshibashiTada2b}.

In summary, we argued that by adapting $L_{0}-(L_{1}+L_{-1})/2$ as the holomorphic part of the Hamiltonian in stead of $L_{0}$, we can derive a %new -- omitted by Ishibashi-san's suggestion 03/24/2015
CFT with the continuous Virasoro algebra. The new theory can be considered to be the infinite circumference limit of the original one. Another viewpoint might be that the continuous spectrum can be attributed to the non-trivial world-sheet metric incorporated by the present procedure. Should we adopt this viewpoint, it would be natural to expect that the present result can be extended to other quantum statistical systems that are susceptible to SSD. While the present analysis uses  the conformal symmetry, it would be interesting to see if the other systems also show continuous spectra \cite{Hotta:2014}.  Of course, CFT's are ubiquitous in the study of string theory, and hence there are many subjects wherein  this new system can be put in good use.

\noindent{\bf Acknowledgements:} 
This study is supported in part by JSPS KAKENHI Grant No. 25400242 and No. 25610066, and the RIKEN iTHES Project. We would like to thank H. Katsura for valuable inputs as well as drawing the authors' attention to Refs. \cite{Doyon:2013paa,Hotta:2014}
. We  would also like to thank M. Asano for useful discussions. T. T  benefitted from the workshop YITP-W-14-4 "Strings and Fields."

\end{document}